**Title:** Single-shot three-dimensional imaging with a metasurface depth camera

**Authors:**

Shane Colburn[1,*] & Arka Majumdar[1,2,*]

**Affiliations:**

[1]Department of Electrical and Computer Engineering, University of Washington, Seattle, Washington 98195, USA.

[2]Department of Physics, University of Washington, Seattle, Washington 98195, USA.

*Correspondence to: scolbur2@uw.edu, arka@uw.edu

**Abstract:**

Depth imaging is vital for many emerging technologies with applications in augmented reality, robotics, gesture detection, and facial recognition. These applications, however, demand compact and low-power systems beyond the capabilities of state-of-the-art depth cameras. Here, we leverage ultrathin dielectric metasurfaces to demonstrate a solution that, with a single surface, replicates the functionality of a high-performance depth camera typically comprising a spatial light modulator, polarizer, and three lenses. Using cylindrical nano-scatterers that can arbitrarily modify the phase of an incident wavefront, our metasurface passively encodes two complementary optical responses to depth information in a scene with a single camera snapshot. By decoding the captured data in software, our system produces a fully reconstructed image and transverse depth map of three-dimensional scenes with a fractional ranging error of 1.7%. We demonstrate the first visible wavelength and polarization-insensitive metasurface depth camera, representing a significant form factor reduction for such systems.

# Introduction:

Conventional cameras capture two-dimensional projections of intensity information from three-dimensional scenes without any knowledge of object depths. While this is often sufficient, depth information is crucial to the operation of numerous next-generation technologies, such as autonomous transportation and gesture recognition in augmented reality. A variety of approaches for collecting depth information from a scene exist[1,2], but these often require active illumination or multiple viewpoints that prohibitively increase system size. Alternatively, there are depth from defocus methods[3–6] that obtain depth information from a sequence of images under different defocus settings; however, this typically requires a dynamic setup where the optics are physically adjusted between each capture. Moreover, information theoretic calculations show that the depth precision from such depth from defocus methods is fundamentally limited for a standard lens[7,8], as the point spread function (PSF) varies slowly with changes in depth and it is often ambiguous whether an object is defocused away from or towards the lens.

There are, however, optical elements with more exotic PSFs compared to that of a standard lens, enabling significantly more precise depth discrimination. A prominent example of this is the double-helix PSF (DH-PSF), which distinguishes depths as it produces two foci that rotate continuously in plane in response to shifting the distance of a point source[8–13]. While single-shot depth imaging with a DH-PSF was demonstrated by analyzing an image's power cepstrum[14], the presence of sidelobes in the PSF limited the reconstructed image quality. The image quality can be improved by capturing an additional reference image, albeit at the cost of not being a single-shot capture[8]. For real-time depth imaging, this entails physically adjusting the optics and repetitively capturing images. In one implementation of a double-helix-based depth camera[13], this functionality was achieved via a spatial light modulator (SLM) whose phase was switched between that of a DH-PSF and a cubic phase mask. This required an extensive setup comprising an imaging lens paired with a polarizer and the SLM, as well as a 4f correlator with two Fourier transform lenses[13].

Metasurfaces present a compelling route for miniaturizing such systems. These elements consist of quasiperiodic arrays of subwavelength scatterers that can alter the phase, amplitude, and polarization of incident light in an ultrathin form factor[15–18], enabling a class of flat lenses[19–28]. While depth imaging was recently reported using a metasurface-based plenoptic camera[29], this required circularly polarized illumination and relied on small aperture (21.65 µm) lenses that significantly limit lateral imaging resolution. Separately, a metasurface-based DH-PSF[30] was shown and used for depth imaging[31], but this operated in the infrared, relied on a separate refractive imaging lens, and did not reconstruct the scene. In this paper, we demonstrate a miniature, visible wavelength depth camera by collapsing the functionality of multiple supplemental lenses into a single polarization-insensitive spatially multiplexed metasurface with an aperture area of 2 mm$^2$. Coupled with deconvolution software, our system generates three-dimensional images, i.e., both a transverse depth map and a monochromatic focused scene image with a single snapshot under incoherent, visible illumination.

**Results:**

Our system comprises a dual aperture metasurface element, consisting of two adjacent metasurfaces with distinct and complementary PSFs that work in tandem to enable simultaneous scene reconstruction and depth acquisition (see Figure 1A). These metasurfaces form two separate and non-overlapping sub-images on a sensor array with a single snapshot. The detected sub-images are then processed via a deconvolution algorithm to produce both a focused image and a corresponding depth map. The functionality of the SLM, 4f correlator, and imaging lens present in the standard implementation[13] are combined into a single surface by setting the phase of each metasurface to a sum of a lens term and a wavefront coding term as

$$\Phi = \Phi_{lens} + \Phi_{WC}, (1)$$

in which

$$\Phi_{lens} = \frac{2\pi}{\lambda}\left(f - \sqrt{x^2 + y^2 + f^2}\right), (2)$$

where $\lambda$ is the optical wavelength, $x$ and $y$ are the in-plane position coordinates, and $f$ is the focal length of the lens. In our design, both metalenses have a 1 mm wide square aperture, focal length $f = 5$ mm, and a design wavelength $\lambda = 532$ nm.

For the first metasurface, the wavefront coding term exhibits a PSF that is highly invariant with depth, whereas the wavefront coding term for the second metasurface generates a double-helix PSF that is highly sensitive to changes in object depth[13]. Both metasurfaces are made of silicon nitride cylindrical nanoposts[26,32–34]. Silicon nitride was selected due to its CMOS compatibility and transparency over the visible wavelength range[35], while cylindrical nanoposts provide the benefit of polarization insensitivity[22]. The nanoposts in our design have a thickness $t = 600\ nm$ and period $p = 400\ nm$. Figure 1B shows their transmission coefficient as a function of diameter calculated by rigorous coupled-wave analysis[36] (see supplementary materials and Figure S1 for transmission coefficient data as a function of lattice constant). Using the simulated transmission coefficient's phase as a lookup table, a diameter is assigned to impart the desired phase for each position in equation (1).

The depth-invariant design is achieved via an extended depth of focus (EDOF) metalens[32] with the wavefront coding term

$$\Phi_{WC} = \frac{\alpha}{L^3}(x^3 + y^3), (3)$$

where $L$ is half the aperture width, and $\alpha$ is a constant that multiplies the cubic phase modulation term to generate an accelerating Airy beam that produces a misfocus-insensitive PSF [13,32,34,37–39]. Figure 2A shows simulated PSFs for the EDOF metalens with $\alpha = 20\pi$ as the point source is shifted to different depths along the optical axis, demonstrating the uniformity in the response. While this metalens does not focus to a point and therefore captures blurry images, by calibration with a single PSF measurement and subsequent deconvolution, focused images can be reconstructed with high fidelity over a wide depth range[32].

Complementing the depth-invariant design, the depth-variant metasurface leverages a DH-PSF. The wavefront coding term of a double-helix metalens is determined via a sum of Laguerre-Gaussian modes[8,40,41] and a block-iterative weighted projections algorithm[42–44] (see supplementary materials and Figure S2). Figure 2B shows simulated PSFs for the designed DH metalens, exhibiting distinct intensity patterns for each depth unlike the case of the EDOF metalens. In an imaging system, the DH metalens creates two spatially shifted and rotated copies of objects, where the rotation angle between the two copies is determined by the distance of the object being imaged.

We then fabricated the dual aperture metasurface to validate our design. Figure 3A shows a picture of the sample mounted on a microscope slide. An optical micrograph of the adjacent metasurfaces in Figure 3B shows the asymmetry in their phase profiles, where the different colored zones correspond to regions of different diameters that were selected to achieve $2\pi$ phase coverage. In Figure 3C and 3D, scanning electron micrographs depict zoomed in views of the nanoposts on a rectangular lattice at normal and $45^o$ incidence respectively.

We then measured the PSFs of the fabricated metasurfaces. As expected, the PSF of the EDOF metasurface varied minimally with depth (Figure 4A), while that of the DH metalens (Figure 4B) strongly depends on the point source distance, demonstrating a large change ($\sim 87^o$) in orientation angle over the measured depth range (Figure 4C). Furthermore, the orientation angle of the lobes in the measured DH-PSFs as a function of depth agrees very well with the theory[8,40] (see supplementary materials). The measured diffraction and transmission efficiencies of the full combined metasurface aperture were 75% and 91% respectively.

Armed with our dual metasurface aperture exhibiting complementary depth responses, we performed a computational imaging experiment on a scene consisting of patterns on standard printer paper located at different depths. Our patterns were illuminated with a wideband incoherent white light source but the light incident on our sensor was spectrally filtered via a 1 nm full width at half maximum bandpass filter centered at 532 nm wavelength. Each captured image comprised two sub-images (Figure 5A). The full scene was then reconstructed by applying a total variation-regularized deconvolution algorithm[45] to the sub-image produced by the EDOF metalens and its measured PSF. After segmenting the reconstructed image and labelling objects for depth estimation, we could estimate the experimental DH-PSF for each object of interest. Figure 5B shows the PSF calculated from the image of a "3" character located 6.5 cm away from the metasurface. With the calculated PSFs, we estimated the depth per object (see Methods for further details). Applying this computational framework, we reconstructed scenes and calculated depth maps for the "3" character of Figure 5B located at 5 different depths in the 6.5 cm to 16.9 cm range (Figure 5D-E).

As the DH-PSF's rotation angle depends on the wavefront's accumulated Gouy phase[40,41], off-axis aberrations such as field curvature induce rotation offsets to the PSF that vary as a function of field angle (i.e., the angle to an object in the scene as measured from the optical axis). In a refractive lens system with multiple surfaces that mitigate aberrations from off-axis light (e.g., Petzval field curvature, coma, etc.), the resulting focal shift and rotation offset are negligible and the depth can be extracted directly from a calibration curve[13] as in Figure 4C. Our metalens, however, does not correct for these off-axis aberrations (Figure S3). Hence, naively treating all fields angles in the same manner produces erroneous depth estimates as the focal shift is nonnegligible. To address this, our algorithm accounts for focal shifts induced by off-axis aberrations and correspondingly compensates the rotation angle to improve the depth estimation accuracy by calculating the additional Gouy phase due to field angle (details of the reconstruction and depth estimation algorithm are provided in the Methods and supplementary materials).

For the case of the single object "3" character at five different depths, the accuracy of our estimation is demonstrated in Figure 5C, where the estimated and true depths strongly agree. In this case, accounting for off-axis aberrations had minimal effect as there was little rotation offset to mitigate because the "3" characters were located near the center of the field of view. We then applied our framework to a scene comprising more than one object located off-axis with higher field angles, consisting of a further located "U" character and a closer "W" character. Here, the captured data (Figure 6A) and the subsequently reconstructed image (Figure 6B) allowed us to estimate distinct double-helix PSFs for each character, shown in Figure 6C and 6D for the "U" and "W" respectively. A naive depth estimation without accounting for off-axis focal shift yields highly erroneous depth estimates; however, once the change in Gouy phase due to the field angle of each character is compensated for, the estimates agree well with the true depths once again (Figure 6E). With the depth estimates of both Figure 5 and Figure 6, our system achieves a fractional ranging error of 1.7%, higher than but of similar order compared to existing commercial passive depth cameras but with a much more compact form factor.

**Discussion:**

While various depth estimation techniques exist, our method enables 3-D imaging of scenes in an ultra-compact form factor without having to take multiple snapshots under different optical configurations. By combining the imaging lens and the wavefront coding steps into a single aperture, the size is reduced significantly, albeit at the cost of introducing off-axis and chromatic aberrations from the metalenses. These aberrations, however, are largely mitigated by limiting the optical bandwidth in detection and accounting for the field angle dependence of the focal length when calculating depths. The form factor reduction will be beneficial for a variety of systems, such as head-mounted displays for augmented reality which impose stringent size limitations on sensors. Shifting the functionality of the SLM and 4f correlator into the dual aperture metalens not only contributed to this size reduction, but also eliminated the time multiplexing required in previously reported PSF engineering methods[13]. Eliminating this time multiplexing serves a dual purpose: it reduces the system complexity and also circumvents the issue of a scene changing between sequential captures. Although the spatial multiplexing of two metasurfaces does induce parallax, the center-to-center separation of each metasurface poses a negligible angular separation (less than 0.4°) for the average object depth in our experiments.

      We demonstrated a compact and visible wavelength depth camera for three-dimensional imaging based on a dual aperture optical metasurface. Our system relies on imparting two complementary wavefront coding functions on light from a scene, enabling simultaneous focused scene reconstruction at all distances and depth discrimination for objects in the scene with a single image snapshot. Compared to existing implementations of depth cameras, we

demonstrated an ultra-compact solution with a 2 mm² optical aperture and without requiring a separate imaging lens, 4f correlator, or spatial light modulator. While use of metasurfaces must contend with off-axis aberrations via computational correction and a limited operating bandwidth, recent works demonstrating achromatic lensing[29,32,46–52] and wide-angle field of view correction by stacking metasurfaces[53] are feasible routes for circumventing these issues. Although in this work we focused on the 5 cm to 35 cm range, applicable to gesture recognition for augmented reality systems, the optical design is readily adaptable to other length scales and operating wavelengths by appropriately tuning the cubic phase strength of the EDOF metalens, aperture size, and focal length.

**Methods:**

*Metasurface Design:* To optimize the phase for the double-helix metalens, a block-iterative weighted projections algorithm[42–44] was used that axially constrained the diffracted intensity along the optical axis at 8 different parallel planes, decomposed the metasurface mask into a linear combination of Laguerre-Gaussian modes, and enforced a phase-only constraint for the mask (see supplementary materials for additional details of the algorithm). The nanopost designs were first simulated using the Stanford S4 rigorous coupled-wave analysis package[36] to extract their transmission coefficients. These coefficients were then assigned to their corresponding diameters and treated as complex amplitude pixels in a custom wave optics MATLAB code to simulate the full designs. The wave optics simulation was based on the angular spectrum method[54].

*Fabrication:* Our process began with a cleaved piece of glass from a 100 mm double side polished fused silica wafer. The silicon nitride layer was first deposited via plasma-enhanced chemical vapor deposition at 350°C. The sample was then spin coated with ZEP 520A and an 8 nm Au/Pd charge dissipation layer was sputtered on top. Both metasurface patterns were subsequently exposed adjacent to one another using a JEOL JBX6300FS electron-beam lithography system at 100kV. After stripping the Au/Pd layer, the sample was developed in amyl acetate. A 50 nm layer of aluminum was evaporated and lifted off via sonication in methylene chloride, acetone, and isopropyl alcohol. The silicon nitride layer was then etched with the remaining aluminum as a hard mask using an inductively coupled plasma etcher with a $CHF_3$ and $SF_6$ chemistry. The remaining aluminum was finally removed by immersing the sample in AD-10 photoresist developer. An Au/Pd layer was sputtered on top of the sample for charge dissipation when capturing scanning electron micrographs.

*Experiment:* To measure the point spread functions, a 50 μm pinhole was aligned with the sample and illuminated from behind with a LED source. For imaging experiments, the pinhole was removed and objects on printer paper were illuminated with a white light LED array panel source. For both the point spread functions and images, the captured signal was limited in bandwidth via a 1 nm full width at half maximum spectral bandpass filter centered at 532 nm wavelength. The images and PSFs were magnified via a custom relay microscope comprising an objective and tube lens. The experimental setups and corresponding part numbers for components used in this work are shown in Figure S4 and S5 in the supplementary information for PSF measurement and imaging respectively. The transmission efficiency was calculated by taking the power ratio of the light on the sensor side of the metasurface to that on the source side. The diffraction efficiency was calculated by taking the ratio of the power at the metasurface on the sensor side to that at the focal plane. These powers were measured by integrating the intensity within the area of the metasurface aperture from images when it was backside illuminated.

*Deconvolution:* The reconstructed scene images are calculated by deconvolving the cubic sub-images using a total variation-regularized deconvolution algorithm. This deconvolution problem is solved using an open source MATLAB library based on the split Bregman method[45], which iteratively solves the reconstruction problem. After segmenting the reconstructed scene and labelling objects for depth estimation, we applied a Kaiser window in each desired subregion of the image with a labelled object. Subsequent deconvolution of the subregions via a Wiener filter applied to the double-helix sub-image provided an estimate of the DH-PSF for each object of interest. With the PSF estimates for each labelled object, the orientation angles of the lobes were extracted and compared against the experimentally calibrated angle response of the DH-PSF as a function of depth (Figure 4C), providing a depth estimate for each object. We calculate the focal shift due to off-axis aberrations by finding ray intersections and determine the subsequent change in Gouy phase and rotation angle by using an ABCD formalism for the Gaussian complex beam parameter (see supplementary materials for details). As scene reconstruction and depth estimation per segmented object average 26.5 and 0.8 seconds respectively using an ordinary personal laptop computer (12 GB RAM, Intel CORE i7) with the algorithm implemented in MATLAB, real-time processing would not be possible, though video data could be processed offline after data capture. Significant speedups to achieve real-time processing are feasible, however, if dedicated hardware were used, such as field-programmable gate arrays (FPGAs) or graphics processing units (GPUs).

**Acknowledgements**: This research was supported by a Samsung GRO grant and the UW Reality Lab, Facebook, Google, and Huawei. Part of this work was conducted at the Washington Nanofabrication Facility / Molecular Analysis Facility, a National Nanotechnology Coordinated Infrastructure (NNCI) site at the University of Washington, which is supported in part by funds

**References:**

1. Cyganek, B. & Siebert, J. P. *An Introduction to 3D Computer Vision Techniques and Algorithms*. (John Wiley & Sons, 2011).
2. Geng, J. Structured-light 3D surface imaging: a tutorial. *Adv. Opt. Photon., AOP* **3**, 128–160 (2011).
3. Schechner, Y. Y. & Kiryati, N. Depth from Defocus vs. Stereo: How Different Really Are They? *International Journal of Computer Vision* **39**, 141–162 (2000).
4. Chaudhuri, S. & Rajagopalan, A. N. *Depth From Defocus: A Real Aperture Imaging Approach*. (Springer Science & Business Media, 2012).
5. Xiong, Y. & Shafer, S. A. Depth from focusing and defocusing. in *Proceedings of IEEE Conference on Computer Vision and Pattern Recognition* 68–73 (1993). doi:10.1109/CVPR.1993.340977
6. Darrell, T. & Wohn, K. Pyramid based depth from focus. in *Proceedings CVPR '88: The Computer Society Conference on Computer Vision and Pattern Recognition* 504–509 (1988). doi:10.1109/CVPR.1988.196282
7. Shechtman, Y., Sahl, S. J., Backer, A. S. & Moerner, W. E. Optimal Point Spread Function Design for 3D Imaging. *Phys. Rev. Lett.* **113**, 133902 (2014).
8. Greengard, A., Schechner, Y. Y. & Piestun, R. Depth from diffracted rotation. *Opt. Lett., OL* **31**, 181–183 (2006).
9. Pavani, S. R. P. *et al.* Three-dimensional, single-molecule fluorescence imaging beyond the diffraction limit by using a double-helix point spread function. *PNAS* **106**, 2995–2999 (2009).

10. Pavani, S. R. P. & Piestun, R. Three dimensional tracking of fluorescent microparticles using a photon-limited double-helix response system. *Opt. Express, OE* **16**, 22048–22057 (2008).
11. Badieirostami, M., Lew, M. D., Thompson, M. A. & Moerner, W. E. Three-dimensional localization precision of the double-helix point spread function versus astigmatism and biplane. *Appl. Phys. Lett.* **97**, 161103 (2010).
12. Thompson, M. A., Lew, M. D., Badieirostami, M. & Moerner, W. E. Localizing and Tracking Single Nanoscale Emitters in Three Dimensions with High Spatiotemporal Resolution Using a Double-Helix Point Spread Function. *Nano Lett.* **10**, 211–218 (2010).
13. Quirin, S. & Piestun, R. Depth estimation and image recovery using broadband, incoherent illumination with engineered point spread functions [Invited]. *Appl. Opt., AO* **52**, A367–A376 (2013).
14. OSA | Single shot three-dimensional imaging using an engineered point spread function. Available at: https://www.osapublishing.org/oe/abstract.cfm?uri=oe-24-6-5946. (Accessed: 13th July 2019)
15. Yu, N. & Capasso, F. Flat optics with designer metasurfaces. *Nat Mater* **13**, 139–150 (2014).
16. Planar Photonics with Metasurfaces | Science. Available at: http://science.sciencemag.org/content/339/6125/1232009. (Accessed: 20th June 2017)
17. Yu, N. *et al.* Light Propagation with Phase Discontinuities: Generalized Laws of Reflection and Refraction. *Science* **334**, 333–337 (2011).
18. Jahani, S. & Jacob, Z. All-dielectric metamaterials. *Nat Nano* **11**, 23–36 (2016).
19. Arbabi, A., Briggs, R. M., Horie, Y., Bagheri, M. & Faraon, A. Efficient dielectric metasurface collimating lenses for mid-infrared quantum cascade lasers. *Opt. Express, OE* **23**, 33310–33317 (2015).
20. West, P. R. *et al.* All-dielectric subwavelength metasurface focusing lens. *Opt. Express, OE* **22**, 26212–26221 (2014).
21. Lu, F., Sedgwick, F. G., Karagodsky, V., Chase, C. & Chang-Hasnain, C. J. Planar high-numerical-aperture low-loss focusing reflectors and lenses using subwavelength high contrast gratings. *Opt. Express, OE* **18**, 12606–12614 (2010).
22. Arbabi, A., Horie, Y., Ball, A. J., Bagheri, M. & Faraon, A. Subwavelength-thick lenses with high numerical apertures and large efficiency based on high-contrast transmitarrays. *Nature Communications* **6**, ncomms8069 (2015).
23. Lin, D., Fan, P., Hasman, E. & Brongersma, M. L. Dielectric gradient metasurface optical elements. *Science* **345**, 298–302 (2014).
24. Fattal, D., Li, J., Peng, Z., Fiorentino, M. & Beausoleil, R. G. Flat dielectric grating reflectors with focusing abilities. *Nat Photon* **4**, 466–470 (2010).
25. Aberration-Free Ultrathin Flat Lenses and Axicons at Telecom Wavelengths Based on Plasmonic Metasurfaces - Nano Letters (ACS Publications). Available at: http://pubs.acs.org/doi/abs/10.1021/nl302516v. (Accessed: 21st June 2017)
26. Low-Contrast Dielectric Metasurface Optics - ACS Photonics (ACS Publications). Available at: http://pubs.acs.org/doi/abs/10.1021/acsphotonics.5b00660. (Accessed: 20th June 2017)
27. Khorasaninejad, M. *et al.* Metalenses at visible wavelengths: Diffraction-limited focusing and subwavelength resolution imaging. *Science* **352**, 1190–1194 (2016).

28. Klemm, A. B. *et al.* Experimental high numerical aperture focusing with high contrast gratings. *Opt. Lett., OL* **38**, 3410–3413 (2013).
29. Lin, R. J. *et al.* Achromatic metalens array for full-colour light-field imaging. *Nature Nanotechnology* **14**, 227 (2019).
30. Jin, C., Zhang, J. & Guo, C. Metasurface integrated with double-helix point spread function and metalens for three-dimensional imaging. *Nanophotonics* **8**, 451–458 (2019).
31. Jin, C. *et al.* Dielectric metasurfaces for distance measurements and three-dimensional imaging. *AP* **1**, 036001 (2019).
32. Colburn, S., Zhan, A. & Majumdar, A. Metasurface optics for full-color computational imaging. *Science Advances* **4**, eaar2114 (2018).
33. Fan, Z.-B. *et al.* Silicon Nitride Metalenses for Close-to-One Numerical Aperture and Wide-Angle Visible Imaging. *Phys. Rev. Applied* **10**, 014005 (2018).
34. Colburn, S. & Majumdar, A. Simultaneous varifocal and broadband achromatic computational imaging using quartic metasurfaces. *arXiv:1904.09622 [physics]* (2019).
35. Colburn, S. *et al.* Broadband transparent and CMOS-compatible flat optics with silicon nitride metasurfaces [Invited]. *Opt. Mater. Express, OME* **8**, 2330–2344 (2018).
36. Liu, V. & Fan, S. S4 : A free electromagnetic solver for layered periodic structures. *Computer Physics Communications* **183**, 2233–2244 (2012).
37. Dowski, E. R. & Cathey, W. T. Extended depth of field through wave-front coding. *Appl. Opt., AO* **34**, 1859–1866 (1995).
38. Cathey, W. T. & Dowski, E. R. New paradigm for imaging systems. *Appl. Opt., AO* **41**, 6080–6092 (2002).
39. Zhan, A., Colburn, S., Dodson, C. M. & Majumdar, A. Metasurface Freeform Nanophotonics. *Scientific Reports* **7**, 1673 (2017).
40. Piestun, R., Schechner, Y. Y. & Shamir, J. Propagation-invariant wave fields with finite energy. *J. Opt. Soc. Am. A, JOSAA* **17**, 294–303 (2000).
41. Schechner, Y. Y., Piestun, R. & Shamir, J. Wave propagation with rotating intensity distributions. *Phys. Rev. E* **54**, R50–R53 (1996).
42. Pavani, S. R. P. & Piestun, R. High-efficiency rotating point spread functions. *Opt. Express, OE* **16**, 3484–3489 (2008).
43. Piestun, R., Spektor, B. & Shamir, J. Wave fields in three dimensions: analysis and synthesis. *J. Opt. Soc. Am. A, JOSAA* **13**, 1837–1848 (1996).
44. Piestun, R. & Shamir, J. Control of wave-front propagation with diffractive elements. *Opt. Lett., OL* **19**, 771–773 (1994).
45. Getreuer, P. Total Variation Deconvolution using Split Bregman. *Image Processing On Line* **2**, 158–174 (2012).
46. Arbabi, E., Arbabi, A., Kamali, S. M., Horie, Y. & Faraon, A. Controlling the sign of chromatic dispersion in diffractive optics with dielectric metasurfaces. *Optica, OPTICA* **4**, 625–632 (2017).
47. Chen, W. T. *et al.* A broadband achromatic metalens for focusing and imaging in the visible. *Nature Nanotechnology* 1 (2018). doi:10.1038/s41565-017-0034-6
48. Wang, S. *et al.* A broadband achromatic metalens in the visible. *Nature Nanotechnology* 1 (2018). doi:10.1038/s41565-017-0052-4
49. Khorasaninejad, M. *et al.* Achromatic Metalens over 60 nm Bandwidth in the Visible and Metalens with Reverse Chromatic Dispersion. *Nano Lett.* **17**, 1819–1824 (2017).

50. Wang, S. *et al.* Broadband achromatic optical metasurface devices. *Nature Communications* **8**, 187 (2017).
51. Shrestha, S., Overvig, A. C., Lu, M., Stein, A. & Yu, N. Broadband achromatic dielectric metalenses. *Light: Science & Applications* **7**, 85 (2018).
52. Chen, W. T., Zhu, A. Y., Sisler, J., Bharwani, Z. & Capasso, F. A broadband achromatic polarization-insensitive metalens consisting of anisotropic nanostructures. *Nature Communications* **10**, 355 (2019).
53. Arbabi, A. *et al.* Miniature optical planar camera based on a wide-angle metasurface doublet corrected for monochromatic aberrations. *Nature Communications* **7**, ncomms13682 (2016).
54. Goodman, J. W. *Introduction to Fourier Optics*. (Roberts and Company Publishers, 2005).

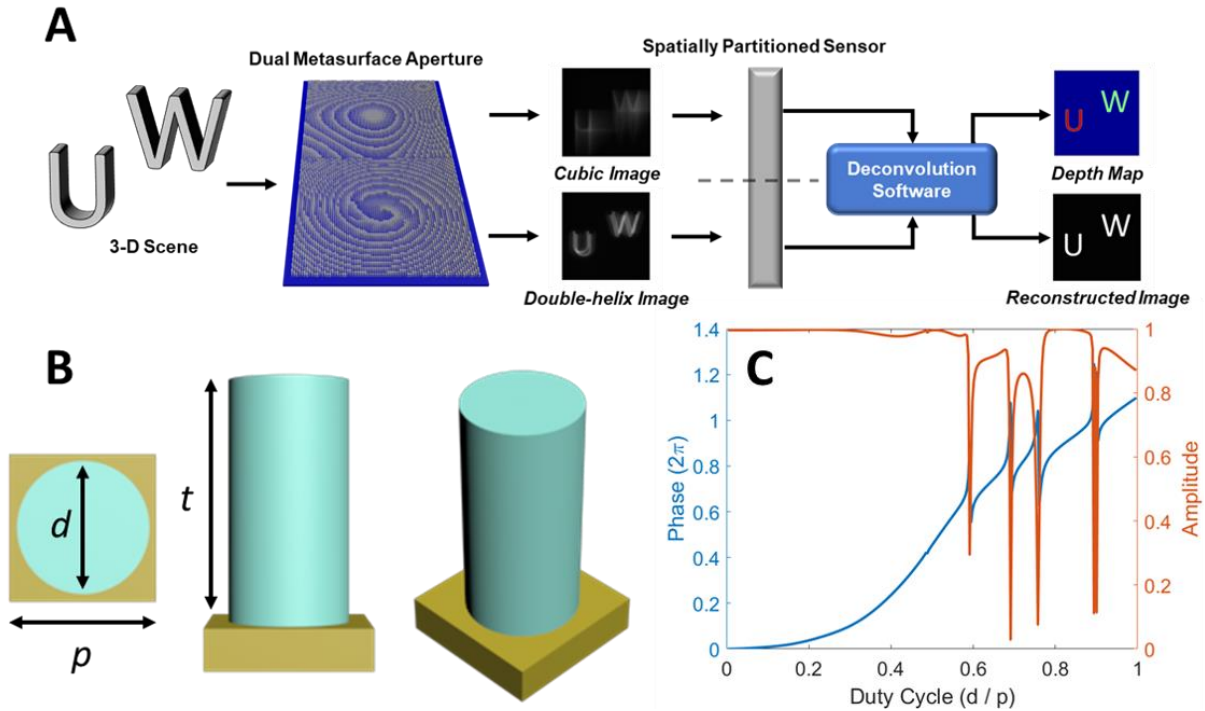

Figure 1: **System Design** (A) Light from a scene incident on the dual aperture metasurface will be captured on a sensor as two side-by-side sub-images: one of them depth-variant and the other one depth-invariant. These sub-images will then be computationally processed to output both a reconstructed scene and a transverse depth map. (B) Schematic of the silicon nitride cylindrical nanoposts on a silicon dioxide substrate. The nanoposts have a lattice constant of $p$, diameter $d$, and thickness $t$. (C) The transmission coefficient (phase and amplitude) as a function of duty cycle for the designed nanoposts. The pillars have a thickness $t = 600\ nm$ and periodicity $p = 400\ nm$.

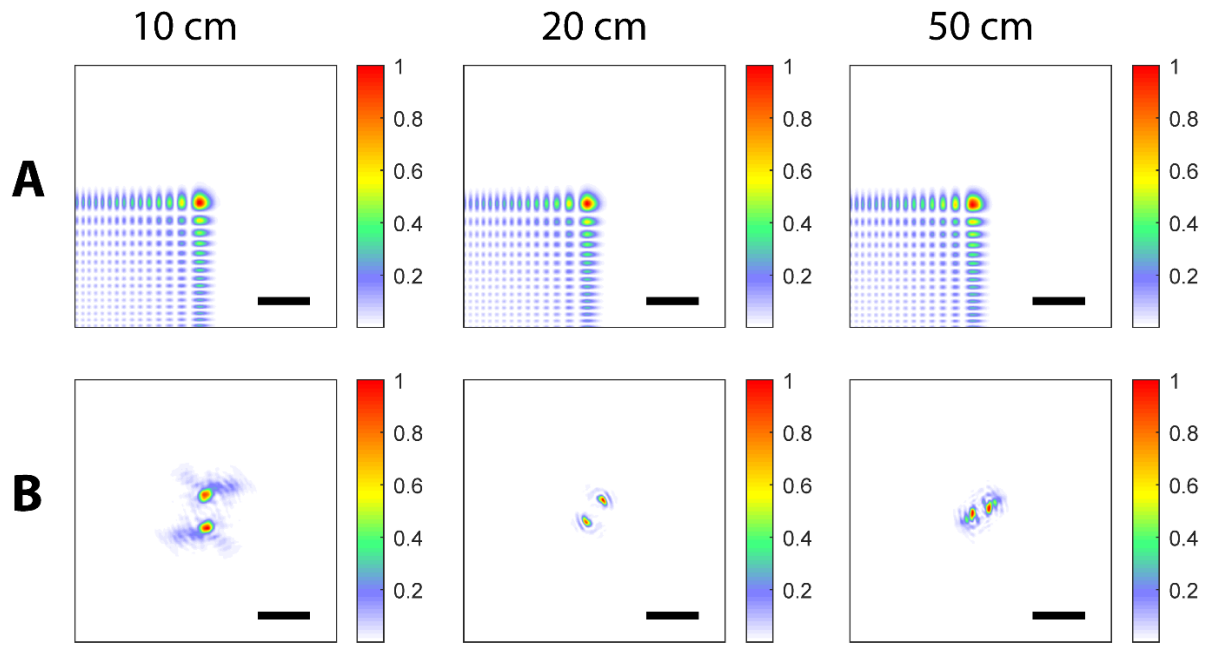

Figure 2: **Simulated metasurface point spread functions:** The normalized intensities of the simulated PSFs are shown for the EDOF (A) and DH-PSF (B) metalenses for three different object distances. Scale bar 32 μm.

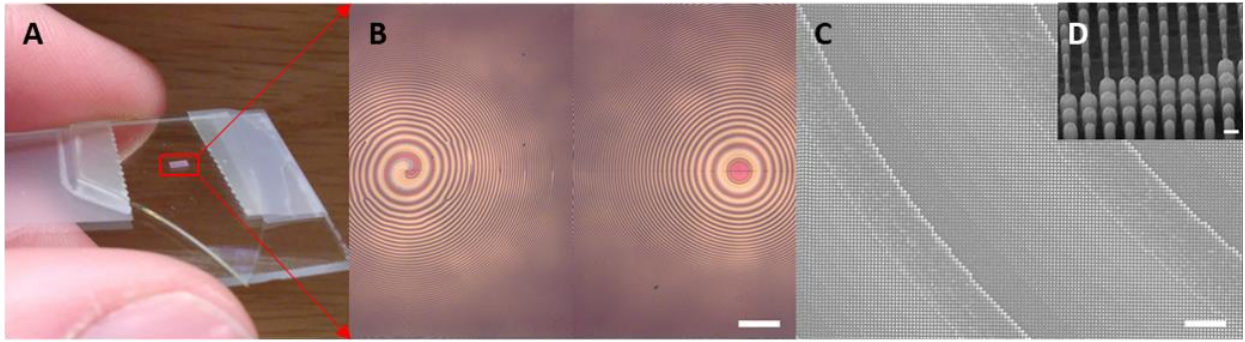

Figure 3: **Fabricated Metasurface** (A) Optical image of the metasurface on a glass slide for testing. (B) Optical microscope image of the dual aperture metasurface. Scale bar 0.125 mm. Scanning electron micrographs at normal (C) and $45^o$ incidence (D) where the scale bars are 5 μm and 300 nm respectively.

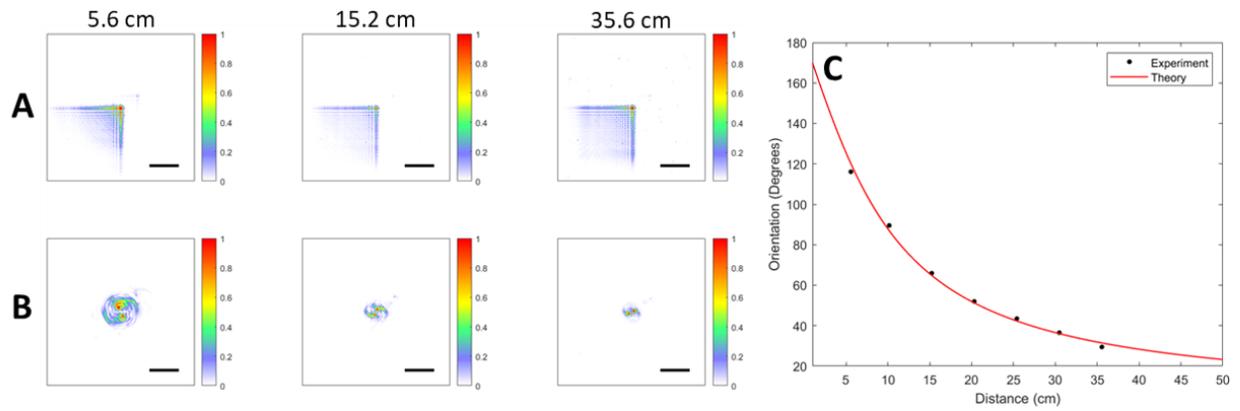

Figure 4: **Metasurface Characterization** Normalized measured intensities of the point spread functions for the EDOF (A) and DH-PSF (B) metalenses for three different object distances. Scale bar 78 μm. (C) Orientation angle of the double-helix foci as a function of object distance.

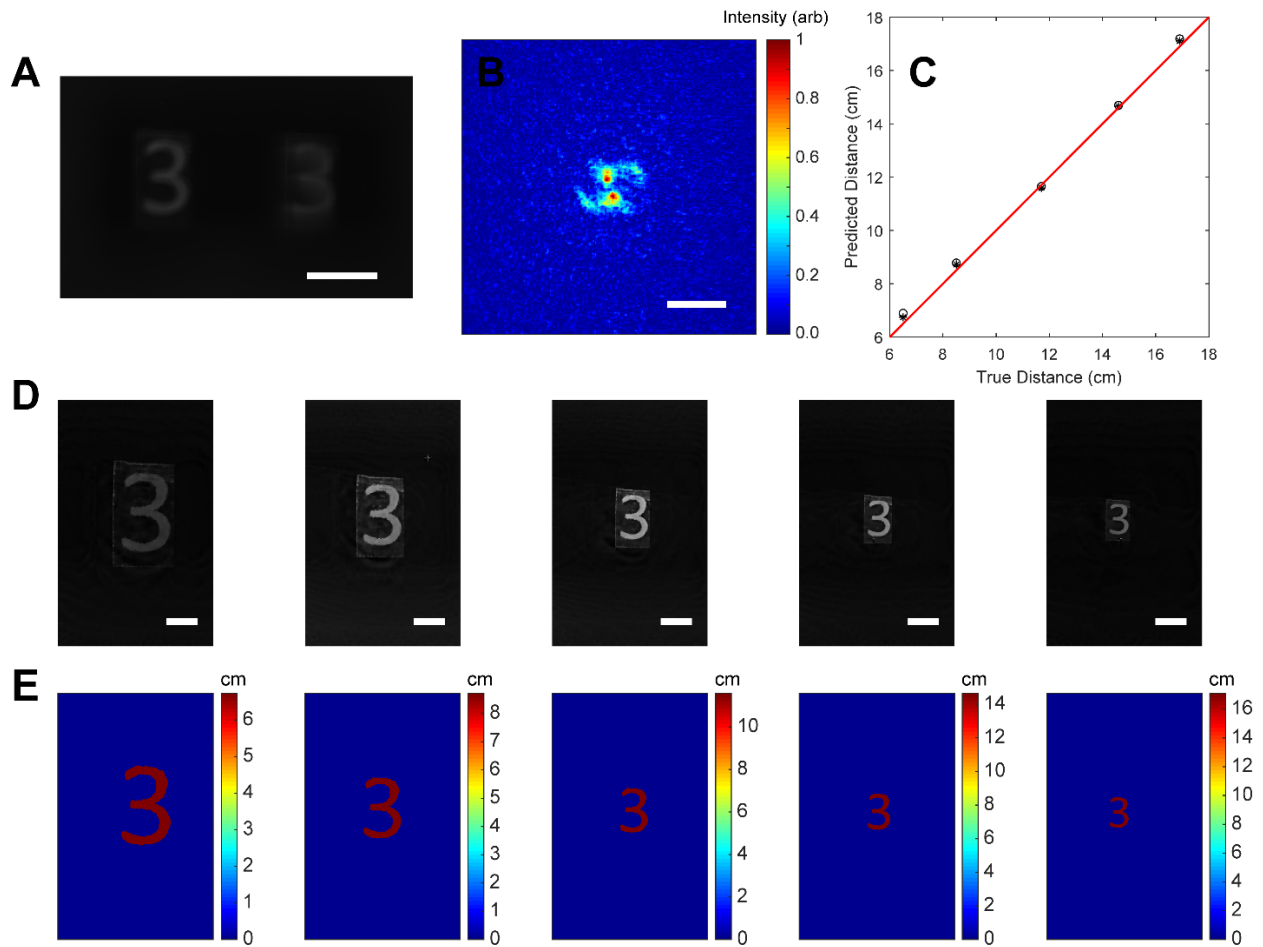

Figure 5: **Single Object Depth Imaging** (A) Raw and unpartitioned image with the double-helix metalens sub-image on the left and the EDOF metalens sub-image on the right. Scale bar 0.5 mm. (B) Estimated DH-PSF from the image in (A). Predicted distances compared to the true distances are plotted in (C) for the case of imaging a "3" character at five different distances, where the reconstructed images and depth maps are shown in (D) and (E) respectively. The circles and asterisks in (C) correspond respectively to depth estimates without and with corrections accounting for changes in Gouy phase due to field angle. The red line denotes the performance of a perfect depth estimation algorithm. Scale bars are 78 μm and 0.2 mm in (B) and (D) respectively.

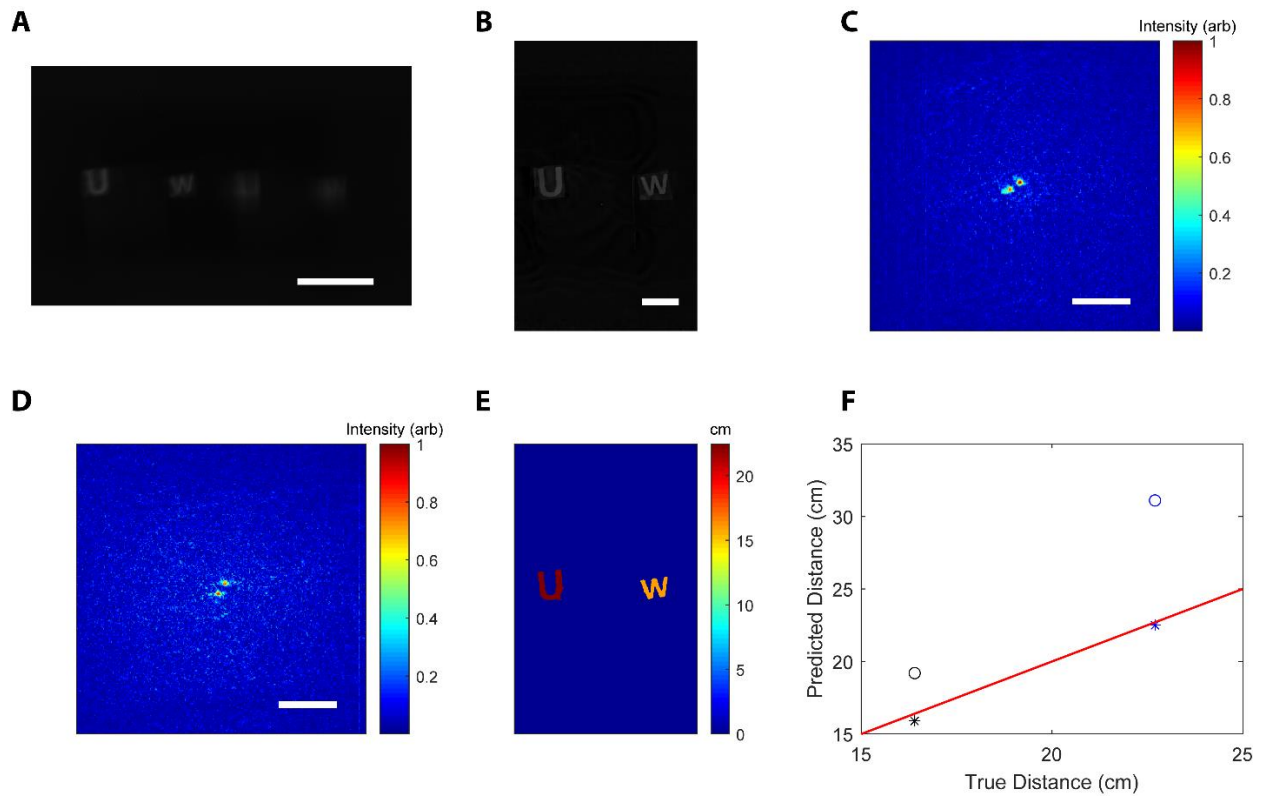

Figure 6: **Imaging Multiple Objects** (A) Raw and unpartitioned image with the double-helix metalens sub-image on the left and the EDOF metalens sub-image on the right of a scene with "U" and "W" characters located at different distances. Scale bar 0.51 mm. (B) Reconstructed object scene with a scale bar of 0.2 mm. Estimated DH-PSFs are shown for the "U" (C) and "W" (D) with scale bars of 78 μm. (E) The calculated transverse depth map for the scene. (F) Predicted distances compared to the true distances are plotted, where the circles and asterisks correspond respectively to depth estimates without and with corrections accounting for changes in Gouy phase due to field angle. The blue and black points correspond to the "U" and "W" characters respectively. The red line denotes the performance of a perfect depth estimation algorithm.